\documentclass[12pt]{article}
\usepackage{amsfonts,amssymb,amsmath}
\usepackage[dvips]{epsfig}
\newcounter{tempeq}
\begin{document}
\title{Generalized $su(1,1)$ coherent states for pseudo harmonic oscillator and their nonclassical properties }
\author{B. Mojaveri\thanks{Email: bmojaveri@azaruniv.ac.ir; bmojaveri@gmail.com}\hspace{2mm} and \hspace{2mm}A. Dehghani\thanks{Email: a\_dehghani@tabrizu.ac.ir,  alireza.dehghani@gmail.com}  \\
{\small {\em Department of Physics, Azarbaijan Shahid Madani University, PO Box 51745-406, Tabriz, Iran\,}}\\
{\small {\em Department of Physics, Payame Noor University, PO Box
19395-4697, Tehran, Iran \,}}} \maketitle
\begin{abstract}
In this paper we define a non-unitary displacement operator, which by
acting on the vacuum state of the pseudo harmonic oscillator (PHO),
generates new class of generalized coherent states (GCSs). An interesting
feature of this approach is that, contrary to the Klauder-Perelomov
and Barut-Girardello approaches, it does not require the existence
of dynamical symmetries associated with the system under consideration. These
states admit a resolution of the identity through positive definite
measures on the complex plane. We have shown that the realization of
these states for different values of the deformation parameters leads
to the well-known Klauder-Perelomov and Barut-Girardello CSs associated with the $su(1,1)$ Lie algebra. This is why we call them
the generalized $su(1,1)$ CSs for the PHO. Finally, study of some statistical characters such as squeezing,
anti-bunching effect and sub-Poissonian statistics reveals
that the constructed GCSs have indeed nonclassical features.
\\
\\
 {\bf Keywords:} Nonlinear Coherent States, Pseudo Harmonic Oscillator, Sub-Poissonian Statistics, Squeezing Effect.
\end{abstract}

\section{Introduction}
Coherent states (CSs), were first established by Schr\"{o}dinger
\cite{Schr¨odinger} as the eigenvectors of the boson annihilation
operator, $\hat{a}$, corresponding to the Heisenbereg-Weyl Lie
algebra. They play an important role in quantum optics and provide
us with a link between quantum and classical oscillators. Moreover,
these states can be produced by acting of the Glauber displacement
operator, $D(z)=e^{z \hat{a}^{\dag}-\overline{z}\hat{a}}$, on the
vacuum states, where $z$ is a complex variable. These states were
later applied successfully to some other models based on their Lie
algebra symmetries by Glauber \cite{Glauber1, Glauber2}, Klauder
\cite{Klauder1, Klauder2}, Sudarshan \cite{Sudarshan}, Barut and
Girardello \cite{Barut} and Perelomov \cite{Perelomov}.
Additionally, for the models with one degree of freedom either
discrete or continuous spectra- with no remark on the existence of a
Lie algebra symmetry- Gazeau et al proposed new CSs, which were
parametrized by two real parameters \cite{Gazeau, Antoine}.
Moreover, there exist some considerations in connection with CSs
corresponding to the shape invariance symmetries \cite{Fukui,
Chenaghlou}. To construct CSs, four main different approaches the
so-called Schr\"{o}dinger, Klauder-Perelomov, Barut-Girardello, and
Gazeau-Klauder methods have been found, so that the second and the
third approaches rely directly on the Lie algebra symmetries and
their corresponding generators. Here, it is necessary to emphasize
that quantum coherence of states nowadays pervade many branches of
physics such as quantum electrodynamics, solid-state physics, and
nuclear and atomic physics, from both theoretical and experimental
viewpoints.

In addition to CSs, squeezed states (SSs) are becoming increasingly
important. These are the non-classical states of the electromagnetic
field in which certain observables exhibit fluctuations less than in
the vacuum state \cite{Stoler}. These states are important because
they can achieve lower quantum noise than the zero-point
fluctuations of the vacuum or coherent states. Over the last four
decades there have been several experimental demonstrations of
nonclassical effects, such as the photon anti-bunching
\cite{Kimble}, sub-Poissonian statistics \cite{Short, Teich}, and
squeezing \cite{Slusher, Wu}. On the other hand, considerable
attention has been paid to the deformation of the harmonic
oscillator algebra of creation and annihilation operators
\cite{Biedenharn}. Some important physical concepts such as the CSs,
the even- and odd-CSs for ordinary harmonic oscillator
have been extended to deformation case. Moreover, there exist
interesting quantum effects, and related quantum states that are
namely superposition states exhibiting quantum interference effects
\cite{Yurke, Noel}. Besides, superpositions of CSs can
be prepared in the motion of a trapped ion \cite{Matos1, Monroe}.
With respect to the nonclassical effects, the coherent states turn
out to define the limit between the classical and nonclassical
behavior.

Another type of generalization of CSs is the nonlinear coherent
states (NLCSs), or f-CSs. The NLCSs, $|z, {f}\rangle$, are
right-hand eigenstates of the product of nonlinear function
$f(\hat{N})$ of the number operator $\hat{N}$ and the boson
annihilation operator $\hat{a}$, i.e. they satisfy
${f}(\hat{N})\hat{a} |z, {f}\rangle=z|z, {f}\rangle$. The nature of
the nonlinearity depends on the choice of the function $f(\hat{N})$
\cite{Manko}. These states may appear as stationary states of the
center-of mass motion of a trapped ion \cite{Matos2, Raffa}. NLCSs
exhibit nonclassical features such as quadrature squeezing,
sub-Poissonian statistics, anti-bunching, self-splitting effects and
so on [27- 34].

The $su(1,1)$ Lie algebra is of great interest in quantum optics
because it can characterize many kinds of quantum optical systems
\cite{Barut, Perelomov, Perelemov1, Vourdas}. It has recently been
used by many researchers to investigate the nonclassical properties
of light in quantum optical systems \cite{Zhang}. In particular, the
bosonic realization of $su(1,1)$ describes the degenerate and
non-degenerate parametric amplifiers \cite{WODKIEWICZ}. The squeezed
states and nonlinear SSs of photons have been considered in terms of
$su(1,1)$ Lie algebra and the CSs associated with this algebra
\cite{Alkader}.

In the present paper, we want to construct new type NLCSs for pseudo
harmonic oscillator (PHO). This exactly solvable quantum model is
the sum of the harmonic oscillator and the inversely quadratic
potential was proposed by Goldman and Krivchenkov
\cite{Krivchenkov}, i.e.
\begin{eqnarray}
&&\hspace{-1.5cm}V(x)=\left[\frac{1}{2}\mu
w^2x^2+\frac{\hbar}{2\mu}\frac{\lambda(\lambda-1)}{x^2}\right],
\end{eqnarray}
where $m, w$ and $\lambda$ respectively represent the mass of the
particle, the frequency and the strength of the external field.
Sometimes this system has been called isotonic potential \cite{Hall,
Hall1, Hall2}. PHO may be more suitable model for the description of
vibrating molecules. For this reason, the study of CSs for PHO is of
great importance which has recently been studied Klauder-Perelomov
and Barut-Girardello type CSs in the framework of $su(1,1)$ Lie
algebra symmetry \cite{Agarwal1,dehghani1, Tavassoly1}.

The aim of this work is introducing a new approach to construct GCSs for
PHO. This approach based on the generalization of the displaced
operator associated with su(1,1) Lie algebra which will be acting on
the vacuum stats of PHO. This approach is extension of the our
previous work in connection to new type NLCSs for harmonic
oscillator associated with the Heisenbereg-Weyl algebra
\cite{dehghani}. An interesting feature of this approach is that,
Contrary to the Klauder-Perelomov and Barut-Girardello approach,
this does not require for existence of dynamical symmetries
associated with the considered system. To construction of such
states, we need only to the raising operator associated with the
considered system in the framework of supersymmetric quantum mechanics. These states admit a resolution of the identity
through positive definite and non-oscillating measures on the
complex plane. We have shown that these states are NLCSs and for
different values of the deformation parameter lead to the well-known
Klauder-Perelomov and Barut-Girardello CSs for PHO. Some interesting
features are found. For instance, we have shown that they evolve in
time as like as the canonical CSs, in other words the constructed
GCSs possess the temporal stability property. Furthermore, it has
been discussed in detail that they have indeed nonclassical features
such as squeezing, anti-bunching effect and sub-Poissonian
statistics, too.

 This paper is organized as follows: in section 2, we briefly
review on a su(1,1) Lie algebra symmetry of PHO and construct the new CSs $|z\rangle_{r}^{\lambda}$, via
generalized analogue of the displacement operators acting on the
vacuum state. In order to realize the resolution of the identity, we
have found the positive definite measures on the complex plane. With
a review on these states, the relation between $su(1,1)$
Klauder-Perelomov and Barut-Girardello CSs of PHO with constructed CSs will be obvious. It has
been shown that these states can be considered as eigenstates of a
certain annihilation operator, then they can be interpreted as NLCSs
with a special nonlinearity function. Furthermore, in section 3 by
evaluating some physical quantities, we discuss their non-classical
properties. Finally, we conclude the paper in section 4.

\section{New GCSs for PHO}

In Refs. \cite{ dehghani1, sasaki, Dodonov}, it has been shown that
the second-order differential operators
\begin{eqnarray}
&&\hspace{-1.5cm} {J}_{\pm}^{\lambda}:=\frac{1}{4}\left[\left(x\mp\frac{d}{dx}\right)^2-\frac{\lambda(\lambda-1)}{x^2}\right],\nonumber\\
&&\hspace{-1.5cm}{J}_{3}^{\lambda}:=\frac{H^{\lambda}}{2}=\frac{1}{4}\left[-\frac{d^2}{dx^2}+x^2+\frac{\lambda(\lambda-1)}{x^2}\right],
\end{eqnarray}
satisfy the standard commutation relations of $su(1,1)$ Lie algebra
as follows
\begin{eqnarray}
&&\hspace{-1.5cm}\left[{J}_{+}^{\lambda},{J}_{-}^{\lambda}\right]=-2{J}_{3}^{\lambda},
\hspace{20mm}\left[{J}_{3}^{\lambda},{J}_{\pm}^{\lambda}\right]=\pm{J}_{\pm}^{\lambda}.
\end{eqnarray}
Here, $H^{\lambda}$ is the PHO or
Calegero-Sutherland Hamiltonian on the half-line $x$, which satisfy
the eigenvalue equation
$H^{\lambda}|n,\lambda\rangle=(2n+\lambda+\frac{1}{2})$ for
$\mu=\hbar=w=1$. In terms of the Fock states, defined by the
associated Laguerre polynomials \cite{Gradshteyn}
$L_{n}^{\alpha}(x)=\frac{1}{n!}x^{-\alpha}e^{x}\left(\frac{d}{dx}\right)^{n}\left(x^{n+\alpha}e^{-x}\right)$
with $Re(\alpha)>-1$,
\begin{eqnarray}
&&\hspace{-1.5cm}\langle x|n,\lambda\rangle=(-1)^n\sqrt{\frac{2\Gamma(n+1)}{\Gamma(n+\lambda+
\frac{1}{2})}}x^{\lambda}e^{-\frac{x^2}{2}}
L_{n}^{\lambda-\frac{1}{2}}(x^2), \hspace{4mm}\lambda
>\frac{-1}{2},
\end{eqnarray}
one can realizes that the infinite dimensional Hilbert space
$\mathcal{H}^{\lambda}:={\mathrm{span}}\{|n,\lambda\rangle\}|_{n=0}^{\infty}$
products the unitary and positive-integer irreps of $su(1,1)$ Lie
algebra as \setcounter{tempeq}{\value{equation}}
\renewcommand\theequation{\arabic{tempeq}\alph{equation}}
\setcounter{equation}{0} \addtocounter{tempeq}{1}
\begin{eqnarray}
&&\hspace{-15mm}{J}_{+}^{\lambda}|n-1,\lambda\rangle=\sqrt{n\left(n+\lambda-\frac{1}{2}\right)}|n,\lambda\rangle,\\
&&\hspace{-15mm}{J}_{-}^{\lambda}|n,\lambda\rangle=\sqrt{n\left(n+\lambda-\frac{1}{2}\right)}|n-1,\lambda\rangle,\\
&&\hspace{-15mm}{J}_{3}^{\lambda}|n,\lambda\rangle=\left(n+\frac{\lambda}{2}+\frac{1}{4}\right)|n,\lambda\rangle.
\end{eqnarray}
It is straightforward that the orthogonality condition of the
associated Laguerre polynomials lead to the following orthogonality
condition of the basis of the Hilbert space $\mathcal{H}^{\lambda}$:
\renewcommand\theequation{\arabic{equation}}
\setcounter{equation}{\value {tempeq}}
\begin{eqnarray}
&&\hspace{-15mm}\langle n, \lambda|m,
\lambda\rangle:=\frac{2n!}{\Gamma(n+\lambda+\frac{1}{2})}
\int_{0}^{\infty}x^{2\lambda}e^{-x^2}L_{n}^{\lambda-\frac{1}{2}}(x^2)
L_{m}^{\lambda-\frac{1}{2}}(x^2)dx=\delta_{nm}.
\end{eqnarray}
It is useful to stress that the two operators ${J}_{+}^{\lambda}$
and ${J}_{-}^{\lambda}$ are Hermitian conjugate of each others with
respect to the inner product (5) and ${J}_{3}^{\lambda}$ is
self-adjoint operator, too.

According to the definition has already been given in Ref.
\cite{dehghani}, the following GCSs for PHO are produced, here, via generalized analogue of the
displacement operators acting on the vacuum state of PHO, $|0,\lambda\rangle$
\begin{eqnarray}
&&\hspace{-15mm}|z\rangle_{r}^{\lambda}:=
{M_{r}^{\lambda}}^{-\frac{1}{2}}(|z|)_{1}F_{r}\left(\left[\lambda-\frac{1}{2}\right]
,\left[\lambda-\frac{1}{2}, \lambda+\frac{1}{2}, ..., \lambda-\frac{3}{2}+r\right],z{J}_{+}^{\lambda}
\right)|0,\lambda\rangle, \hspace{5mm} r\geq1,
\end{eqnarray}
where $ _{p}F_{q} (. . .)$ is the generalized hypergeometric
function and $z$$(= |z|e^{i\varphi})$ and $r$ are respectively the
coherence and the deformation parameters, respectively. Now, we
show that the well known $su(1,1)$ Klauder-Perelomov and
Barut-Girardello CSs can be considered as a special case of
introduced GCSs. Clearly, $|z\rangle_{r}^{\lambda}$ becomes the
$su(1,1)$ Klauder-Perelomov CSs for the PHO,
$|z\rangle^{\lambda}_{KP}$ \cite{dehghani1,Tavassoly1}), when $r$
tends to unity and $z$ be replaced with $\frac{z}{|z|}\tanh(|z|)$.
Using the series form of the hypergeometric functions and applying
the laddering relations, Eqs. (5), $|z\rangle_{r}^{\lambda}$ can be
expanded into the basis $|n, k\rangle$ as
\begin{eqnarray}
&&\hspace{-15mm}|z\rangle_{r}^{\lambda}={M_{r}^{\lambda}}^{-\frac{1}{2}}(|z|)
\sum_{n=0}^{\infty}{z^{n}\prod_{k=1}^{r-1}{\frac{\Gamma(\lambda+k-\frac{1}{2})}{\Gamma(n+\lambda+k-\frac{1}{2})}}}
\sqrt{\frac{\Gamma(n+\lambda+\frac{1}{2})}{\Gamma(\lambda+\frac{1}{2})\Gamma(n+1)}}|n,\lambda\rangle,
\hspace{5mm} r\geq2,
\end{eqnarray}
where ${M_{r}^{\lambda}}(|z|)$ is chosen so that
$|z\rangle_{r}^{\lambda}$ is normalized, i.e. $^{\lambda}_{r}\langle
z|z\rangle_{r}^{\lambda}=1$, then
\begin{eqnarray}
&&\hspace{-15mm}{M_{r}^{\lambda}}(|z|)=_{1}F_{2r-2}\left(\left[\lambda+\frac{1}{2}\right],
\left[\lambda+\frac{1}{2} , \lambda+\frac{1}{2}, ..., \lambda-\frac{3}{2}+r, \lambda-\frac{3}{2}+r\right],|z|^2\right).
\end{eqnarray}
Now, one can check that for the case $r= 2$, we have
\begin{eqnarray}
&&\hspace{-15mm}|z\rangle_{2}^{\lambda}={M_{2}^{\lambda}}^{-\frac{1}{2}}(|z|)_{1}F_{2}
\left(\left[\lambda-\frac{1}{2}\right],\left[\lambda-\frac{1}{2},\lambda+\frac{1}{2}\right],z{J}_{+}^{\lambda}\right)|0,\lambda\rangle\nonumber\\
&&\hspace{-5mm}={M_{2}^{\lambda}}^{-\frac{1}{2}}(|z|)
\sum_{n=0}^{\infty}{z^{n}\sqrt{\frac{\Gamma(\lambda+\frac{1}{2})}{\Gamma(n+1)
\Gamma(n+\lambda+\frac{1}{2})}}}|n,\lambda\rangle,\nonumber
\end{eqnarray}
meanwhile it satisfies following eigenvalue equation
\begin{eqnarray}
&&\hspace{-15mm}{J}_{-}^{\lambda}|z\rangle_{2}^{\lambda}=z|z\rangle_{2}^{\lambda}.\nonumber
\end{eqnarray}
Then, $|z\rangle_{r\rightarrow 2}^{\lambda}$ is reduced to the
$su(1,1)$ coherency of the Barut-Girardello type, has already given
in Ref \cite{dehghani1, Tavassoly1} corresponding to the PHO model. Also, it should be noticed that, these
states can be categorized as special class of {\em {Generalized
Hypergeometric CSs}} \cite{Appl} have already been made
by Appl et al.

Using the inner product (6), the overlapping of the GCSs
can be calculated as follows:
\begin{eqnarray}
&&\hspace{-15mm}^{\lambda}_{r}\langle z_{1}|z_{2}\rangle_{r}^{\lambda}=
\frac{_{1}F_{2r-2}\left(\left[\lambda+\frac{1}{2}\right],\left[\lambda+\frac{1}{2}, \lambda+\frac{1}{2}, ...,
\lambda-\frac{3}{2}+r, \lambda-\frac{3}{2}+r\right],\overline{z}_{1}z_{2}\right)}{\sqrt{{M_{r}^{\lambda}}(|z_{1}|){M_{r}^{\lambda}}(|z_{2}|)}},\\
&&\hspace{-15mm}^{\lambda}_{r_{1}}\langle
z|z\rangle_{r_{2}}^{\lambda}=
\frac{_{1}F_{r_{1}+r_{2}-2}\left(\left[\lambda+\frac{1}{2}\right],\left[\lambda+\frac{1}{2},
\lambda+\frac{1}{2}, ..., \lambda-\frac{3}{2}+r_{1},
\lambda-\frac{3}{2}+r_{2}\right],|z|^2\right)}{\sqrt{{M_{r_{1}}^{\lambda}}(|z|){M_{r_{2}}^{\lambda}}(|z|)}},\end{eqnarray}
and result that two different kinds of these states are
non-orthogonal, if $r_{1}\neq r_{2}, z_{1}\neq z_{2}$. Now, we are
in a position to introduce the appropriate measure $d\mu_{r}(|z|) :=
K^{\lambda}_{r}(|z|) \frac{d|z|^{2}d\varphi}{2}$ so that the
resolution of the identity is realized for the GCSs
$|z\rangle_{r}^{\lambda}$ in the Hilbert space
$\mathcal{H}^{\lambda}$:
\begin{eqnarray}
&&\hspace{-15mm}1_{\mathcal{H}^{\lambda}} = \oint_{\mathbb{C}(z)}{|z\rangle_{r}^{\lambda}\hspace{1mm}{^{\lambda}_{r}\langle z|}}d\mu_{r}(|z|)\nonumber\\
&&\hspace{-15mm}=2\pi\sum_{n=0}^{\infty}{\left[\prod_{k=1}^{r-1}{\frac{\Gamma(\lambda+k-\frac{1}{2})}{\Gamma(n+\lambda+k-\frac{1}{2})}}\right]^{2}
{\frac{\Gamma(n+\lambda+\frac{1}{2})}{\Gamma(\lambda+\frac{1}{2})\Gamma(n+1)}}|n,\lambda\rangle\hspace{1mm}\langle
n,\lambda|\int_{0}^{\infty}{|z|^{2n+1}\frac{K^{\lambda}_{r}(|z|)}{M_{r}^{\lambda}(|z|)}}}d|z|.
\end{eqnarray}
It is found that using by the integral relation for the Meijers
G-functions (see $\frac{7-811}{4}$ in \cite{Gradshteyn}), these
states resolve the unity operator for any $r$ and $\lambda$ through
a positive definite and non-oscillating measure
\begin{eqnarray}
&&\hspace{-15mm}K^{\lambda}_{r}(|z|)=\frac{\Gamma(\lambda+\frac{1}{2})_{1}F_{2r-2}\left(\left[\lambda+\frac{1}{2}\right],
\left[\lambda+\frac{1}{2}, , \lambda+\frac{1}{2}, ...,
\lambda-\frac{3}{2}+r,
\lambda-\frac{3}{2}+r\right],|z|^2\right)}{\pi
\left[\prod_{k=1}^{r-1}\Gamma(\lambda+k-\frac{1}{2})\right]^{2}}\nonumber\\
&&\hspace{4mm}\times
G^{2r-1,\,\,1}_{2,\,\,2r}\left(|z|^2\,\,\big|^{0,\hspace{0.5mm}
\lambda-\frac{1}{2}}_{0,\hspace{0.5mm}\lambda-\frac{1}{2},\,\lambda-\frac{1}{2},\,\,...,\,
\lambda+r-\frac{5}{2},\,\lambda+r-\frac{5}{2},\hspace{0.5mm}0}\right).
\end{eqnarray}
For $(\lambda; r)=(\frac{3}{4}; 1)$ as well as $(\lambda;
r)=(\frac{3}{2}; 2, 3$ and 4) we have plotted the changes of the
non-oscillating and positive definite measures
$K^{\lambda}_{r}(|z|)$ in terms of $|z|^2$ in Figures 1(a) and 1(b),
respectively.\\\\
\begin{figure}
\centering
\includegraphics[width=445 pt]{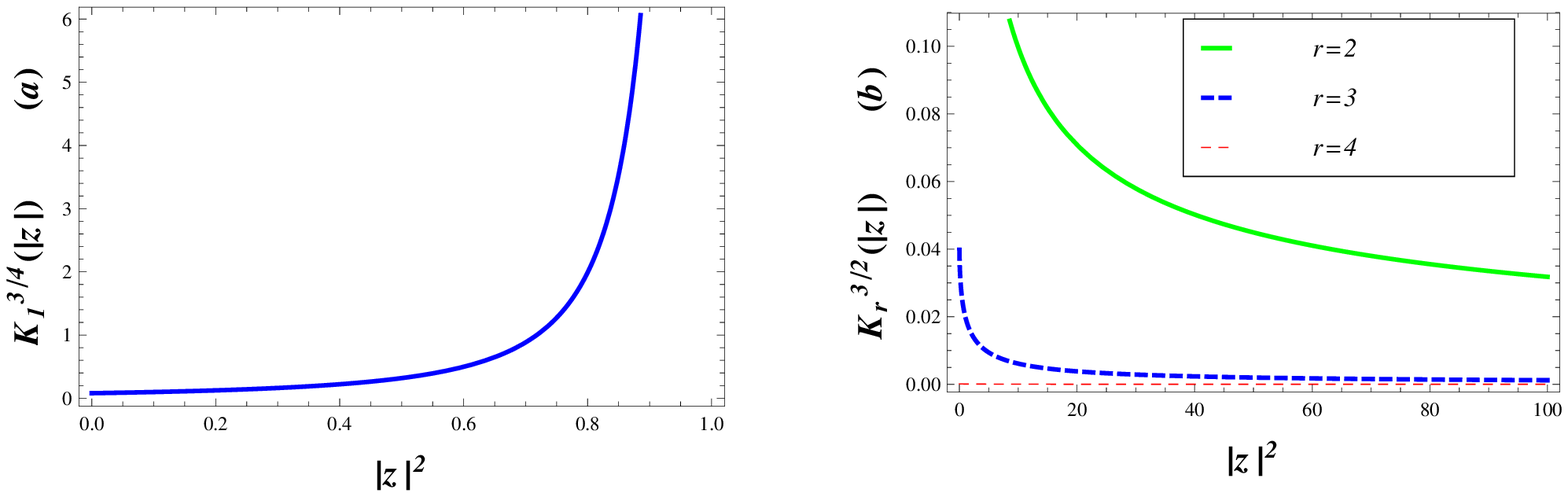}
\caption{} \label{Fig1}\itemize{}
\item Figure. 1. {Plots of the positive definite measures $K^{\lambda}_{r}(|z|)$ in terms of
$|z|^2$; for $\lambda=\frac{3}{4}$ and $r=1$ in $(a)$ likewise
$\lambda=\frac{3}{2}$ and $r(=2,3$ and 4) in $(b)$}.
\end{figure}

$\diamondsuit${\em{Coordinate Representation of} $|z\rangle_{r}^{\lambda}$}\\
Based on a new expression of the Laguerre polynomials as an
operator-valued function given in \cite{Cessa}
\begin{eqnarray}
&&\hspace{-15mm}L_{n}^{\alpha}(y) =
\frac{1}{n!}y^{-\alpha}\left(\frac{d}{dy}-1\right)^n
y^{n+\alpha},
\end{eqnarray}
also, according to Eqs. (4) and (8) we have
\begin{eqnarray}
&&\hspace{-15mm}\langle x|z\rangle_{r}^{\lambda}=\sqrt{\frac{2}{{{M_{r}^{\lambda}}(|z|)\Gamma(\lambda+\frac{1}{2})}}}
\sum_{n=0}^{\infty}{(-z)^{n}\prod_{k=1}^{r-1}{\frac{\Gamma(\lambda+k-\frac{1}{2})}{\Gamma(n+\lambda+k-\frac{1}{2})}}}
x^{\lambda}e^{-\frac{x^2}{2}}L_{n}^{\lambda-\frac{1}{2}}(x^2)\nonumber\\
&&\hspace{-15mm}=\sqrt{\frac{2}{{{M_{r}^{\lambda}}(|z|)\Gamma(\lambda+\frac{1}{2})}}}
\sum_{n=0}^{\infty}{\frac{(-z)^{n}}{n!}\prod_{k=1}^{r-1}{\frac{\Gamma(\lambda+k-\frac{1}{2})}{\Gamma(n+\lambda+k-\frac{1}{2})}}}
e^{-\frac{y}{2}}y^{\frac{1-\lambda}{2}}\left(\frac{d}{dy}-1\right)^n
y^{n+\lambda-\frac{1}{2}}|_{y=x^2}.
\end{eqnarray}
Along with substitution
\begin{eqnarray}
&&\hspace{-15mm} \left(\frac{d}{dy}-1\right)^n
y^{n}=\left(y\frac{d}{dy}+n-y\right)...\left(y\frac{d}{dy}+1-y\right)=\left(y\frac{d}{dy}-y+1\right)_{n},
\end{eqnarray}
it becomes
\begin{eqnarray}
&&\hspace{-17mm}\langle x|z\rangle_{r}^{\lambda}=\left[\frac{2e^{-{y}}y^{{1-\lambda}}}{{{M_{r}^{\lambda}}(|z|)\Gamma(\lambda+\frac{1}{2})}}\right]^{\frac{1}{2}}
\sum_{n=0}^{\infty}{\frac{\left(y\frac{d}{dy}-y+1\right)_{n}}{\prod_{k=1}^{r-1}{\left(\lambda+k-\frac{1}{2}\right)_{n}}}}
\frac{(-z)^{n}}{n!}y^{\lambda-\frac{1}{2}}|_{y=x^2}\nonumber\\
&&\hspace{-17mm}=\left[\frac{2e^{-{y}}y^{{1-\lambda}}}{{{M_{r}^{\lambda}}(|z|)\Gamma(\lambda+\frac{1}{2})}}\right]^{\frac{1}{2}}
{_{1}F_{r-1}\left(\left[y\frac{d}{dy}-y+1\right],\left[\lambda+\frac{1}{2},
...,\lambda-\frac{3}{2}+r\right],-z\right)}y^{\lambda-\frac{1}{2}}|_{y=x^2},
\end{eqnarray}
where $(\alpha)_{_{n}}= \frac{\Gamma(\alpha+n)}{\Gamma(\alpha)}$,
denotes the Pochhammer symbol. For instance, the explicit compact
forms of $|z\rangle_{2}^{\lambda}$ and $|z\rangle_{3}^{\lambda}$ are
\setcounter{tempeq}{\value{equation}}
\renewcommand\theequation{\arabic{tempeq}\alph{equation}}
\setcounter{equation}{0} \addtocounter{tempeq}{1}
\begin{eqnarray}
&&\hspace{-35mm}\langle
x|z\rangle_{2}^{\lambda}=\left[{\left(-\frac{z}{|z|}\right)^{\frac{1}{2}-\lambda}\frac{2x}
{I_{\lambda-\frac{1}{2}}(2|z|)}}\right]e^{-z-\frac{x^2}{2}}J_{\lambda-\frac{1}{2}}(2ix\sqrt{z}),\\
&&\hspace{-35mm}\langle x|z\rangle_{3}^{\lambda}=\frac{1}{\sqrt{_{1}F_{4}\left(\left[\lambda+\frac{1}{2}\right],
\left[\lambda+\frac{1}{2},\lambda+\frac{1}{2},\lambda+\frac{3}{2},\lambda+\frac{3}{2}
\right],|z|^2\right)}}\nonumber\\
&&\hspace{-25mm}\times \sqrt{2}e^{-{x^2}}x^{2(1-\lambda)}\,\,_{1}F_{2}\left(\left[y\frac{d}{dy}-y+1\right],\left[\lambda+\frac{1}{2}, \lambda+\frac{3}{2}\right],-z\right)y^{\lambda-\frac{1}{2}}|_{y=x^2},
\end{eqnarray}
in which $I_{\lambda-\frac{1}{2}}(x)$ is the modified Bessel
function of first type and $J_{\lambda-\frac{1}{2}}(x)$ is the
ordinary Bessel function defined as
$J_{\lambda-\frac{1}{2}}(x)=\sum_{m=0}^{\infty}\frac{(-1)^m(\frac{x}{2})^{\lambda+2m-\frac{1}{2}}}
{m!\Gamma(\lambda+m+\frac{1}{2})}$ \cite{Gradshteyn}.\\
$\diamondsuit${\em{Time Evolution of generalized su(1,1) CSs}}\\
Due to the relations (2) and (5c), we have
\renewcommand\theequation{\arabic{equation}}
\setcounter{equation}{\value {tempeq}}
\begin{eqnarray}
&&\hspace{-14mm}H^{\lambda}|n,\lambda\rangle=(2n+\lambda+\frac{1}{2})|n,\lambda\rangle.
\end{eqnarray}
Then the CSs (8) evolve in time as
\begin{eqnarray}
&&\hspace{-14mm}e^{-itH^{\lambda}}\left|z\right\rangle_{r}^{\lambda}=\frac{e^{-it\left(\lambda+\frac{1}{2}\right)}}{\sqrt{M_{r}^{\lambda}(|z|)}}
\sum_{n=0}^{\infty}{\left(ze^{-i2t}\right)^{n}\prod_{k=1}^{r-1}{\frac{\Gamma(\lambda+k-\frac{1}{2})}{\Gamma(n+\lambda+k-\frac{1}{2})}}}
\sqrt{\frac{\Gamma(n+\lambda+\frac{1}{2})}{\Gamma(\lambda+\frac{1}{2})\Gamma(n+1)}}|n,\lambda\rangle\nonumber\\
&&\hspace{6mm}=e^{-it\left(\lambda+\frac{1}{2}\right)}\left|ze^{-i2t}\right\rangle_{r}^{\lambda},
\end{eqnarray}
and confirm that  $|z\rangle_{r}^{\lambda}$ are temporally stable.
\section{Non-classical properties of $\left|z\right\rangle_{r}^{\lambda}$}
In this section, we will set up detailed studies on statistical
properties of constructed GNCS. For this reason,
proportional nonlinear function associated to them are introduced.
Moreover, to analyze their statistical behavior, some of the
characters including the second-order correlation function, Mandel's
parameter and quadrature squeezing are evaluated. It should be
mentioned that squeezing or antibunching (negativity of Mandel
parameter) are sufficient (unnecessary) for a state to belong to
nonclassical
states \cite{Shchukin}.\\\\
$\diamondsuit${\em{\textbf{Nonlinearity function}}}\\
The question we pose now is whether the $su(1,1)$ CSs
constructed above can be defined as the eigenstates of the
non-Hermitian and deformed annihilation operator
$f(\hat{N}){J}_{-}^{\lambda}$, i.e.
\begin{eqnarray}
&&\hspace{-14mm}f(\hat{N}){J}_{-}^{\lambda}\left|z\right\rangle_{r}^{\lambda}=z\left|z\right\rangle_{r}^{\lambda},
\end{eqnarray}
where $f(\hat{N})$, is determined in terms of the number
operator\footnote{$\hat{N}|n,\lambda\rangle=n|n,\lambda\rangle$.}
$\hat{N}$( $={J}_{3}^{\lambda}-\frac{\lambda}{2}-\frac{1}{4}$),
plays an important role as a nonlinearity function \cite{Matos2}.
Combining definition of the generalized $su(1,1)$ CSs (8) and
laddering relations (5), we get
\begin{eqnarray}
&&\hspace{-14mm}\left[\frac{\Gamma(\hat{N}+\lambda+r-\frac{1}{2})}{\Gamma(\hat{N}+\lambda+\frac{3}{2})}\right]{J}_{-}^{\lambda}
\left|z\right\rangle_{r}^{\lambda}=z\left|z\right\rangle_{r}^{\lambda}.
\end{eqnarray}
So $|z\rangle_{r}^{\lambda}$ can be identified as new classes of
$su(1,1)$ NCSs \cite{Wang1}, with characterized nonlinearity
functions,
$\frac{\Gamma(\hat{N}+\lambda+r-\frac{1}{2})}{\Gamma(\hat{N}+\lambda+\frac{3}{2})}$.
Obviously $f(\hat{N})\longrightarrow1$ when $r\longrightarrow2$. \\\\
$\diamondsuit${\em{\textbf{su(1,1) squeezing}}}\\
We introduce two generalized Hermitian quadrature operators $X_{1}$
and $X_{2}$
\begin{eqnarray}
&&\hspace{-14mm}X_{1}^{\lambda}=\frac{{J}_{+}^{\lambda}+{J}_{-}^{\lambda}}{2},
\hspace{4mm}X_{2}^{\lambda}=\frac{{J}_{-}^{\lambda}-{J}_{+}^{\lambda}}{2i},
\end{eqnarray}
with the commutation relation $[X_{1}^{\lambda}, X_{2}^{\lambda}] =
i{J}_{3}^{\lambda}$. From this communication relation the
uncertainty relation for the variances of the quadrature operators
$X_{i}$ follows
\begin{eqnarray}
&&\hspace{-14mm}\langle(\Delta X_{1}^{\lambda})^2\rangle
\langle(\Delta X_{2}^{\lambda})^2\rangle \geq
\frac{|\langle{J}_{3}^{\lambda}\rangle|^{2}}{4},
\end{eqnarray}
where $\langle(\Delta X_{1}^{\lambda})^2\rangle= \langle
\left(X_{1}^{\lambda}\right)^2\rangle-{\langle
X_{1}^{\lambda}\rangle}^2$ and the angular brackets denote averaging
over an arbitrary normalizable state for which the mean values are
well defined, $\langle X_{i}\rangle={^{\lambda}_{r}\langle
z|}X_{i}|z\rangle_{r}^{\lambda}$. Following Walls (1983) as well as
Wodkiewicz and Eberly (1985) \cite{Walls, WODKIEWICZ} we will say
that the state is $su(1,1)$ squeezed if the condition
\begin{eqnarray}
&&\hspace{-14mm}\langle(\Delta X_{i}^{\lambda})^2\rangle <
\frac{|\langle{J}_{3}^{\lambda} \rangle|}{2},\hspace{4mm}
for\hspace{2mm} i=1 \hspace{2mm}or \hspace{2mm}2,
\end{eqnarray}
is fulfilled. In other words, a set of quantum states are called
SSs if they have less uncertainty in one quadrature
($X_{1}$ or $X_{2}$) than CSs. To measure the degree of
the $su(1,1)$ squeezing we introduce the squeezing factor
$S^{\lambda}_{i}$\cite{Buzek}
\begin{eqnarray}
&&\hspace{-14mm}S^{\lambda}_{i}=\frac{\langle(\Delta
X_{i}^{\lambda})^2\rangle-\frac{|\langle{J}_{3}^{\lambda}
\rangle|}{2}}{\frac{|\langle{J}_{3}^{\lambda} \rangle|}{2}},
\end{eqnarray}
it leads that the $su(1,1)$ squeezing condition takes on the simple
form $S^{\lambda}_{i}< 0$, however maximally squeezing is obtained
for $S^{\lambda}_{i}=-1$. By using of the mean values of the
generators of the $su(1,1)$ Lie algebra, one can derive that
uncertainty in the quadrature operators $X_{i}$ can be expressed as
the following forms
\begin{eqnarray}
&&\hspace{-14mm}\langle(\Delta X_{1(2)}^{\lambda})^2\rangle
=\frac{2\left\langle{J}_{+}^{\lambda}{J}_{-}^{\lambda}\right\rangle+2\left\langle{J}_{3}^{\lambda}\right\rangle\pm\left\langle{{J}_{+}^{\lambda}}^{2}+{{J}_{-}^{\lambda}}^{2}\right\rangle-
{\left\langle{{J}_{-}^{\lambda}}\pm{{J}_{+}^{\lambda}}\right\rangle}^{2}}{4}.\end{eqnarray}
where we have the relations \setcounter{tempeq}{\value{equation}}
\renewcommand\theequation{\arabic{tempeq}\alph{equation}}
\setcounter{equation}{0} \addtocounter{tempeq}{1}
\begin{eqnarray}
&&\hspace{-20mm}\left\langle{J}_{+}^{\lambda}\right\rangle=\overline{\left\langle{J}_{-}^{\lambda}\right\rangle}=
\frac{\Gamma(\lambda+\frac{3}{2})}{\Gamma(\lambda+r-\frac{1}{2})}
\overline{z}\nonumber\\
&&\hspace{-18mm}\times\frac{{_{2}F_{2r-1}\left(\left[\lambda+\frac{1}{2},\hspace{0.5mm}\lambda+\frac{3}{2}\right],
\left[\lambda+\frac{1}{2},\lambda+\frac{1}{2}, ...,
\lambda+r-\frac{3}{2}, \lambda+r-\frac{3}{2},
\lambda+r-\frac{1}{2}\right],|z|^{2}\right)}}{_{1}
F_{2r-2}\left(\left[\lambda+\frac{1}{2}\right],\left[\lambda+\frac{1}{2}
, \lambda+\frac{1}{2}, ..., \lambda+r-\frac{3}{2},
\lambda+r-\frac{3}{2}\right],|z|^2\right)},\nonumber\\
&&\hspace{-20mm}\left\langle{{J}_{+}^{\lambda}}^{2}\right\rangle=
\overline{\left\langle{{J}_{-}^{\lambda}}^{2}\right\rangle}=\frac{{\Gamma(\lambda+\frac{3}{2})\Gamma(\lambda+\frac{5}{2})}}
{{\Gamma(\lambda+r-\frac{1}{2})\Gamma(\lambda+r+\frac{1}{2})}}
\overline{z}^{2}\nonumber\\
&&\hspace{-18mm}\times\frac{{_{1}F_{2r-2}\left(\left[\lambda+\frac{5}{2}\right],\left[\lambda+\frac{1}{2},
\lambda+\frac{3}{2},..., \lambda+r-\frac{1}{2}, \lambda+r+\frac{1}{2}\right],|z|^{2}\right)}}
{_{1}F_{2r-2}\left(\left[\lambda+\frac{1}{2}\right],\left[\lambda+\frac{1}{2} , \lambda+\frac{1}{2}, ...,
\lambda+r-\frac{3}{2}, \lambda+r-\frac{3}{2}\right],|z|^2\right)},\nonumber\\
&&\hspace{-20mm}\left\langle{{J}_{+}^{\lambda}}{J}_{-}^{\lambda}\right\rangle=\frac{{\Gamma(\lambda+\frac{3}{2})^{2}}}
{{\Gamma(\lambda+r-\frac{1}{2})^{2}}}
{|z|}^{2}\nonumber\\
&&\hspace{-18mm}\times\frac{{_{3}F_{2r}\left(\left[\lambda+\frac{1}{2},\hspace{0.5mm}\lambda+\frac{3}{2},\hspace{0.5mm}\lambda+\frac{3}{2}\right]
,\left[\lambda+\frac{1}{2},\hspace{0.5mm}\lambda+\frac{1}{2}...,
\lambda+r-\frac{1}{2},
\lambda+r-\frac{1}{2}\right],|z|^{2}\right)}}{_{1}F_{2r-2}\left(\left[\lambda+\frac{1}{2}\right],
\left[\lambda+\frac{1}{2} , \lambda+\frac{1}{2}, ..., \lambda+r-\frac{3}{2}, \lambda+r-\frac{3}{2}\right],|z|^2\right)},\nonumber\\
&&\hspace{-20mm}\left\langle{J}_{3}^{\lambda}\right\rangle=\left(\lambda+\frac{1}{2}\right)\nonumber\\
&&\hspace{-18mm}\times\frac{{_{2}F_{2r-1}\left(\left[\frac{\lambda}{2}+\frac{5}{4},\hspace{0.5mm}\lambda+\frac{1}{2}\right],\left[\frac{\lambda}{2}+\frac{1}{4},
\lambda+\frac{1}{2}, \lambda+\frac{3}{2}, \lambda+\frac{3}{2},...,
\lambda+r-\frac{3}{2}, \lambda+r-\frac{3}{2}
\right],|z|^{2}\right)}}{_{1}F_{2r-2}\left(\left[\lambda+\frac{1}{2}\right],\left[\lambda+\frac{1}{2}
, \lambda+\frac{1}{2}, ..., \lambda+r-\frac{3}{2},
\lambda+r-\frac{3}{2}\right],|z|^2\right)}.\nonumber
\end{eqnarray}
\begin{figure}
\centering
\includegraphics[width=445 pt]{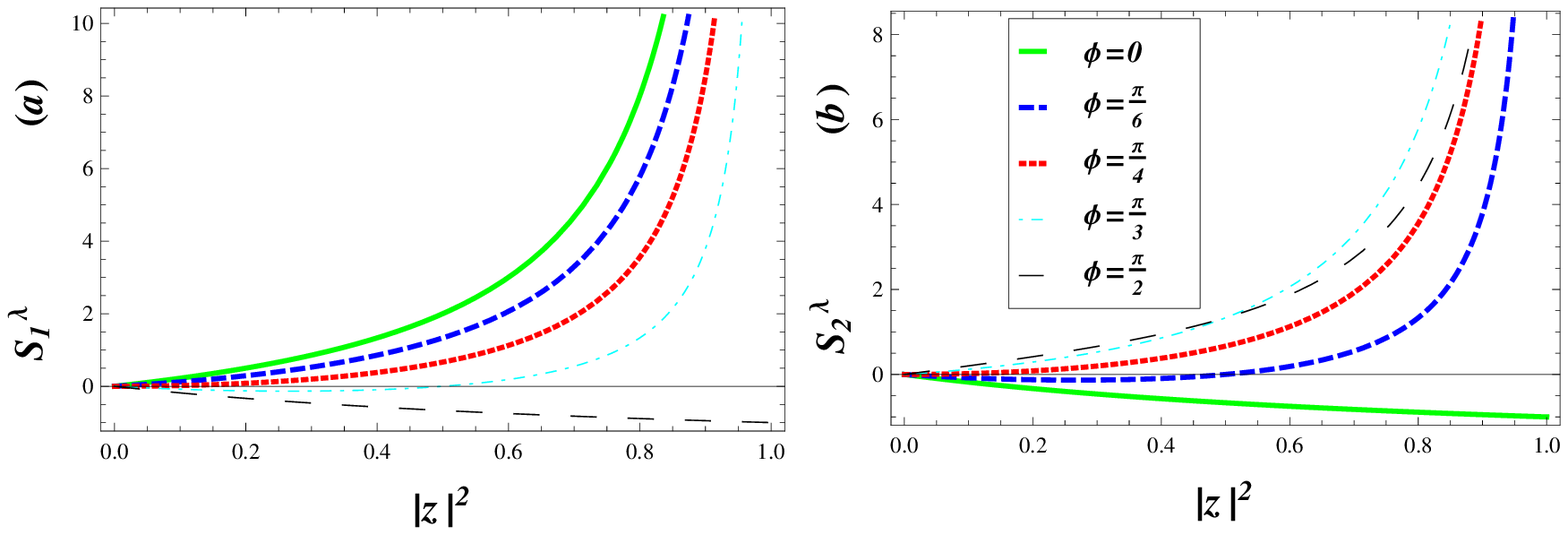}
\caption{} \label{Fig2}\itemize{}
\item Figure. 2. $(a)$ and $(b)$ illustrate squeezing in the $ X_{1}^{\lambda}$
and $ X_{2}^{\lambda}$ quadratures, respectively, against $|z|^2$
for $r=1$ and different values of $\phi$.
\end{figure}

They result that, $\langle(\Delta X_{1(2)}^{\lambda})^2\rangle$ as well as $S^{\lambda}_{1(2)}$, for
any value of $\lambda$, are efficiently dependent on the complex variable $z$($= |z| e^{i\varphi}$ )
and the deformation parameter $r$.\\\\
${\blacksquare}$ Case $r=1$:\\Our calculations show that the squeezing factor $S^{\lambda}_{1}$ is really
independent of $\lambda$. It illustrates that squeezing properties in the $X_{1}$ quadrature arise when
$\varphi$ is increased and culminates where $\varphi$ reaches $\frac{\pi}{2}$ as well as $|z|\rightarrow 1$
(see figure 2(a)). However, figure 2(b) implies that squeezing properties in the $X_{2}$ quadrature is
considerable where $\varphi$ is decreased. It becomes maximal, $S^{\lambda}_{2}\rightarrow -1$, if $\varphi$
tends to zero \cite{Buzek, Tavassoly1}.\\
${\blacksquare}$ Case $r=2$:\\Clearly, for the case $r=2$ we would
not expect to take squeezing neither
in $X_{1}$ nor in $X_{2}$ quadratures \cite{Buzek, Tavassoly1}.\\
${\blacksquare}$ Case $r\geq 3$:\\

\begin{figure}
\centering
\includegraphics[width=300 pt]{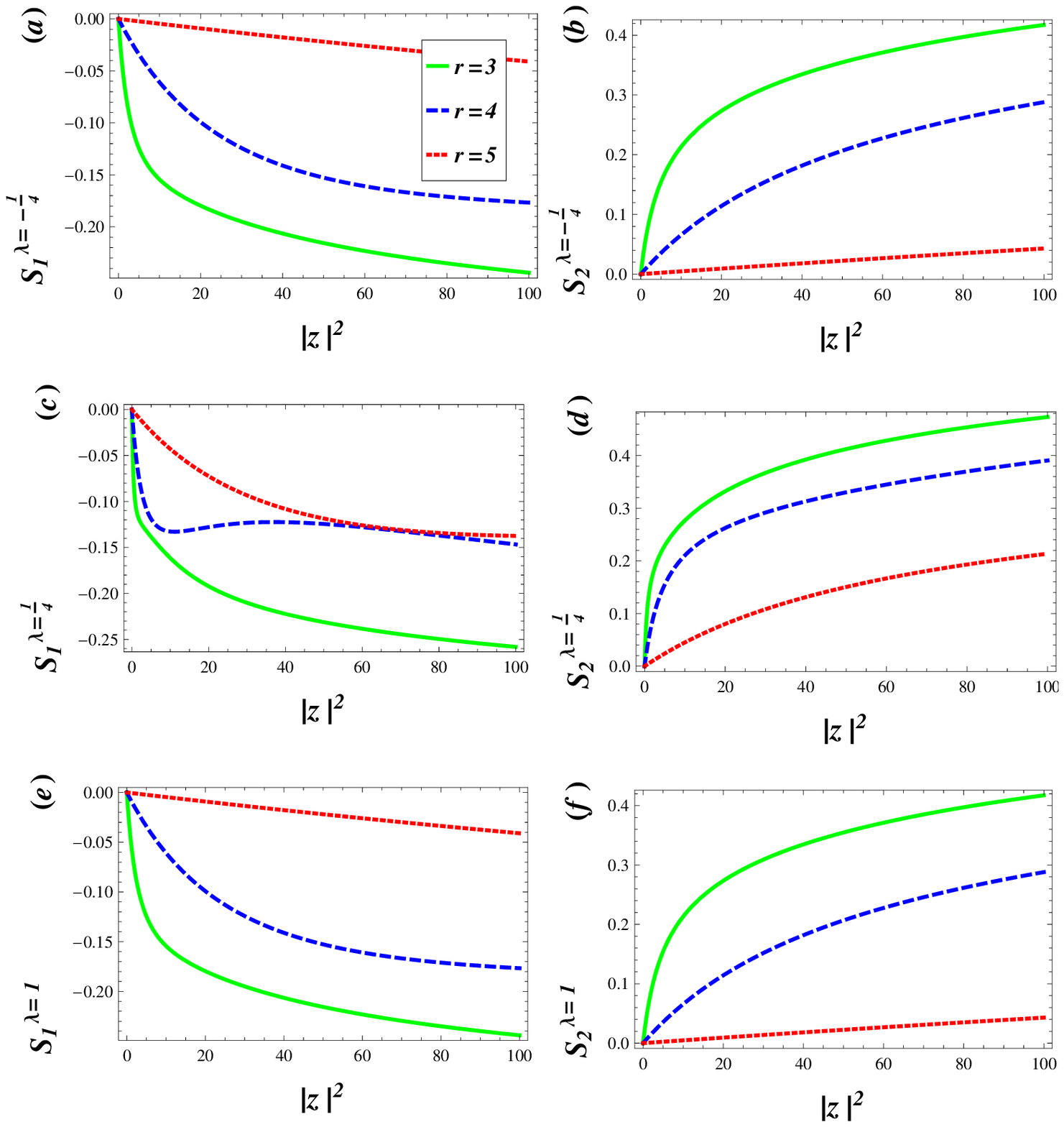}
\caption{} \label{Fig3}\itemize{}
\item Figure. 3. Graphs of uncertainty in the field quadratures $ X_{1}^{\lambda}$ ($a,c,e$) and $ X_{2}^{\lambda}$ $(b,d,f)$, respectively versus $|z|^2$
for different values of $r$ as well as different $\lambda$ while we
choose the phase $\varphi=0$.
\end{figure}
In Figures $3$ we plot the squeezing parameter $S^{\lambda}_{1}$ and
$S^{\lambda}_{2}$ as a function of $|z|^{2}$ for different values of
$r$(= 3, 4 and 5) as well as $\lambda$$( =-\frac{1}{4}, \frac{1}{4},
1)$, here we choose the phase $\varphi=0$. According to Figures
3$(a,c,e)$, it is visible, the squeezing parameter $S^{\lambda}_{1}$
is less than zero for the all regions of $|z|^2$. This indicate that
for $\varphi=0$, the quadrature squeezing occur only in the $X_1$
component for the all regions of $|z|^2$. Also, we show in Figures
$4$ the squeezing parameter $S^{\lambda}_{1}$ and $S^{\lambda}_{2}$
for different values of $\varphi$(= 0, $\frac{\pi}{6}$,
$\frac{\pi}{4}$, $\frac{\pi}{3}$, and $\frac{\pi}{2}$) for fixed
$r=4$. From Figures 3$(a,c,e)$, and Figures 3$(b,d,f)$ we find that
while quadrature squeezing in the $X_1$ component occur for $\phi=0$
and $\phi=\frac{\pi}{6}$, there is quadrature squeezing effect in
the $X_2$ component for $\phi=\frac{\pi}{3}$ and
$\phi=\frac{\pi}{2}$.
\begin{figure}
\centering
\includegraphics[width=300 pt]{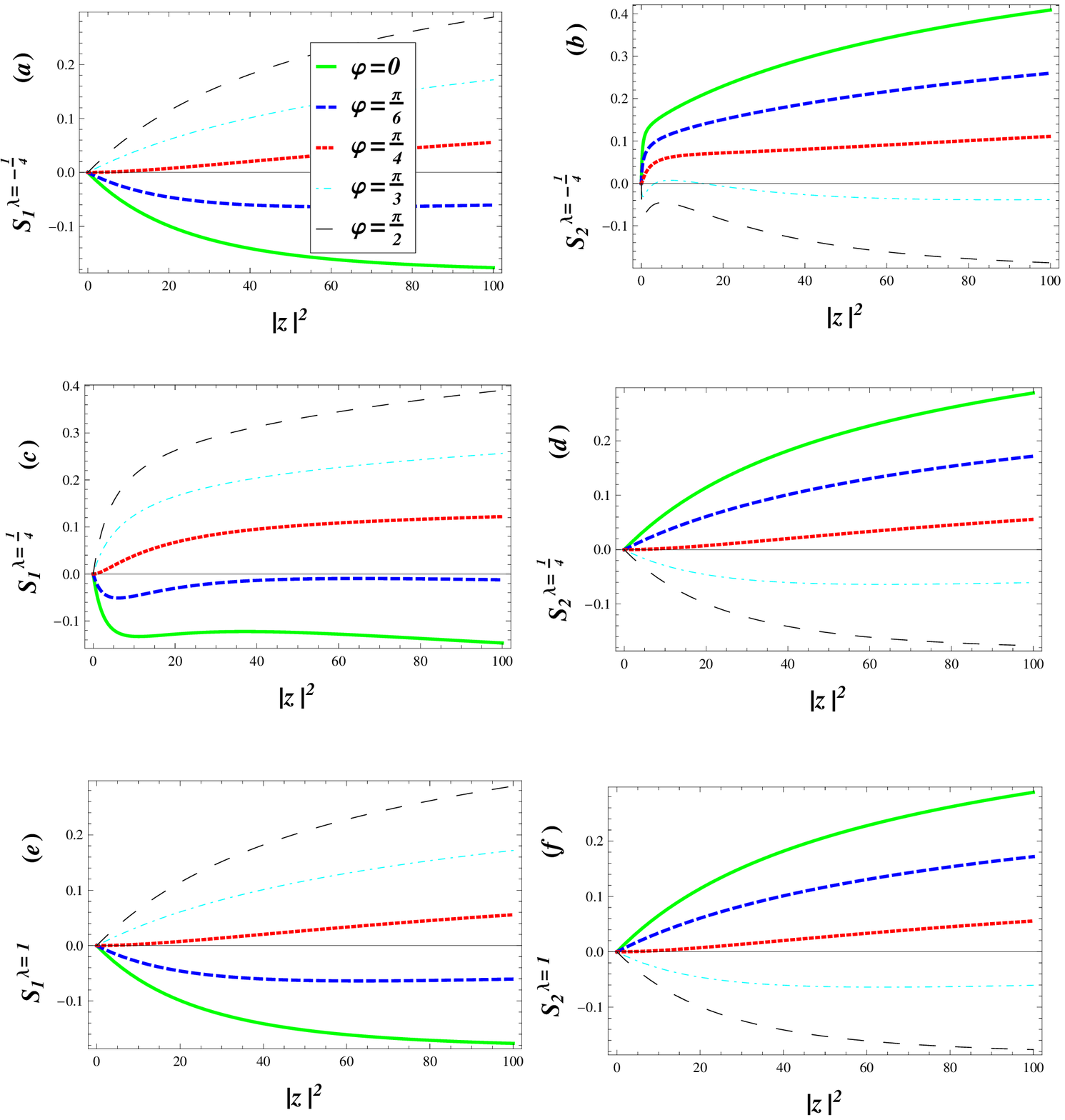}
\caption{} \label{Fig4}\itemize{}
\item Figure. 4. Squeezing in the $ X_{1}^{\lambda}$ and $ X_{2}^{\lambda}$
quadratures have been shown in figures ($a,c,e$) and $(b,d,f)$
respectively, against for $r=4$ and different values of $\varphi$.
\end{figure}\\

$\diamondsuit${\em{\textbf{Anti-bunching effect and sub-Poissonian statistics}}}\\
Now we are in a position to study the anti-bunching effect as well
as the statistics of $|z\rangle_{r}^{\lambda}$ given by equation
(8). We introduce the second-order correlation function for these
states
\renewcommand\theequation{\arabic{equation}}
\setcounter{equation}{\value {tempeq}}
\begin{eqnarray}
&&\hspace{-14mm}\left(g^{(2)}\right)_{r}^{\lambda}(|z|^2)=
\frac{\langle {\hat{N}}^2\rangle_{r}^{\lambda}-{\langle{\hat{N}
\rangle}_{r}^{\lambda}}}{{\langle{\hat{N}
\rangle}_{r}^{\lambda}}^{2}},\end{eqnarray} furthermore, the
inherent statistical properties of the $|z\rangle_{r}^{\lambda}$
follows also from calculating the Mandel parameter
$Q_{r}^{\lambda}(|z|^2)$\footnote{A state for which
$Q_{r}^{\lambda}(|z|^2)> 0$ (or
$\left(g^{(2)}\right)_{r}^{\lambda}(|z|^2)> 1$) is called
super-Poissonian (bunching effect), if $Q = 0$ (or $g^{(2)}= 1$) the
state is called Poissonian, while a state for which $Q< 0$ (or
$g^{(2)}< 1$) is, also, called sub-Poissonian (antibunching
effect).}
\begin{eqnarray}
&&\hspace{-14mm}Q_{r}^{\lambda}(|z|^2)= {\langle{\hat{N}
\rangle}_{r}^{\lambda}}\left[\left(g^{(2)}\right)_{r}^{\lambda}(|z|^2)-1\right].
\end{eqnarray}
In order to find the function
$\left(g^{(2)}\right)_{r}^{\lambda}(|z|^2)$, also Mandel parameter
$Q_{r}^{\lambda}(|z|^2)$, let us begin with the expectation values
of the number operator $\hat{N}$ and of the square of the number
operator $\hat{N}^2$ in the basis of the Fock space states $\left|n,
\lambda\right\rangle$ \setcounter{tempeq}{\value{equation}}
\renewcommand\theequation{\arabic{tempeq}\alph{equation}}
\setcounter{equation}{0} \addtocounter{tempeq}{1}
\begin{eqnarray}
&&\hspace{-12mm}{\langle{\hat{N}
\rangle}_{r}^{\lambda}}=\frac{|z|^2}{\lambda+\frac{1}{2}}\left(\frac{\Gamma(\lambda+\frac{5}{2})}{\Gamma(\lambda+r+\frac{1}{2})}\right)^{2}\nonumber\\
&&\hspace{-5mm}\times\frac{{_{1}F_{2r-2}\left(\left[\lambda+\frac{3}{2}\right],\left[\lambda+\frac{3}{2},
\lambda+\frac{3}{2},...,
\lambda+r-\frac{1}{2},\lambda+r-\frac{1}{2}\right],|z|^{2}\right)}}{_{1}F_{2r-2}\left(\left[\lambda+\frac{1}{2}\right],
\left[\lambda+\frac{1}{2}, \lambda+\frac{1}{2}, ..., \lambda+r-\frac{3}{2}, \lambda+r-\frac{3}{2}\right],|z|^2\right)},\nonumber\\
&&\hspace{-12mm}\langle{\hat{N}}^2\rangle_{r}^{\lambda}=\frac{|z|^2}{\lambda+\frac{1}{2}}\left(\frac{\Gamma(\lambda+\frac{3}{2})}
{\Gamma(\lambda+r-\frac{1}{2})}\right)^{2}\nonumber\\
&&\hspace{-5mm}\times\frac{{_{2}F_{2r-1}\left(\left[2,
\lambda+\frac{3}{2}\right],\left[1, \lambda+\frac{3}{2},
\lambda+\frac{3}{2},...,\lambda+r-\frac{1}{2},\lambda+r-\frac{1}{2}\right],|z|^{2}\right)}}{_{1}F_{2r-2}\left(\left[\lambda+\frac{1}{2}\right]
,\left[\lambda+\frac{1}{2}, \lambda+\frac{1}{2},
...,\lambda+r-\frac{3}{2},\lambda+r-\frac{3}{2}\right],|z|^2\right)}.\nonumber
\end{eqnarray}
For the case $r=1$, the second-order correlation function can be
calculated to be taken as
$\left(g^{(2)}\right)_{r=1}^{\lambda}(|z|^2)=1+\frac{1}{1+\frac{\lambda}{2}}>1$.
This guaranties that $\left|z\right\rangle_{1}^{\lambda}$ exhibits a
fully bunching effect, or super-Poissonian statistics \cite{Buzek}.
But this situation is changed when $r$ and $\lambda$ are enhanced.
\begin{figure}
\centering
\includegraphics[width=300 pt]{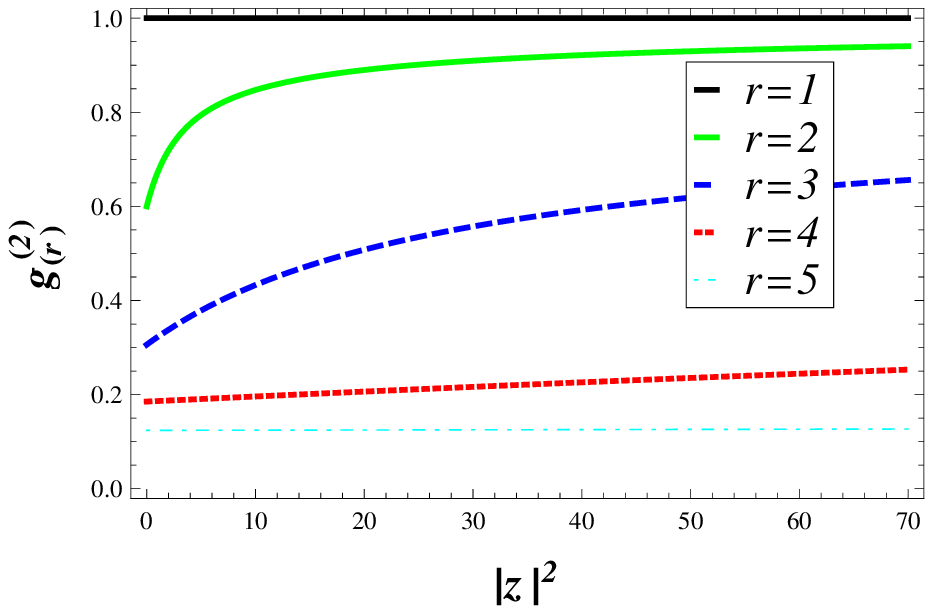}
\caption{} \label{Fig5}\itemize{}
\item Figure. 5. Plot of the correlation function $\left(g^{(2)}\right)_{r}^{\lambda=0}(|z|^2)$
versus $|z|^2$ for different values of $r$. The solid curve,
$\left(g^{(2)}\right)_{r}=1$ corresponds to the canonical CSs.
\end{figure}
In Figure 5, $\left(g^{(2)}\right)_{r}^{\lambda}(|z|^2)$ has been
plotted in terms of $|z|^2$ for several values of $r$(= 2, 3, 4 and
5). From this Figure, it is observed that this parameter is less
than one, for all regions and so the statistics of the
$|z\rangle_{r}^{\lambda}$ is fully sub-Poissonian. Recalling that
each of the nonclassicality indicators is sufficient (not necessary)
for a state to be nonclassical, we observed that the generalized
$su(1,1)$ CSs are indeed nonclassical states.
\section{Conclusions}
Based on a new approach, broad range of generalized $su(1,1)$ CSs
for PHO are constructed. These states realize a resolution of the
identity with positive measures on the complex plane. Non-classical
properties of such states have been reviewed in detail. It has been
shown that they have squeezing properties and follow the
sub-Poissonian statistics. For these reasons the constructed
generalized $su(1,1)$ CSs can be termed as nonclassical states.
Generally, the approach presented here provides a unified method to
construct all relavant CSs introduced in different ways (the
Klauder-Perelomov and Barut-Girardello CSs). The advantage of this
approach is that, one need only the raising operators associated
with the systems under consideration without addressing the
dynamical symmetry of system. Also, this approach can be used to
construct new type CSs for exactly solvable models in the framework
of the quantum mechanics in which the laddering operators are
dependent to the quantum modes, such as the Hydrogen atom and the
Morse model.

\end{document}